# Information-Dense Nonlinear Photonic Physical Unclonable Function

Brian C. Grubel, Bryan T. Bosworth, Michael R. Kossey, A. Brinton Cooper, Mark A. Foster & Amy C. Foster
Department of Electrical and Computer Engineering
Johns Hopkins University
Baltimore, Maryland 21218, USA

*Abstract*—We present a comprehensive investigation into the complexity of a new private key storage apparatus: a novel silicon photonic physical unclonable function (PUF) based on ultrafast nonlinear optical interactions in a chaotic silicon microcavity that is both unclonable and impossible to emulate. This device provides remarkable improvements to total information content (raw cryptographic material), information density, and key generation rates over existing optical scattering PUFs and is also more easily integrated with both CMOS electronics and telecommunications hardware. Our device exploits the natural nonlinear optical behavior of silicon to neutralize commonly used attacks against PUFs and vastly enhance device complexity. We confirm this phenomenon with thorough experimental results on prototype devices and present a detailed estimate of their total information content. Our compact, micron-scale approach represents an entirely new generation of ultrafast and high information density photonic PUF devices that can be directly incorporated into integrated circuits to ensure authenticity and provide secure physical storage of private key material.

I. INTRODUCTION

The globalization of the semiconductor manufacturing industry provides abundant opportunities to attack microelectronic supply chains [1]. Leveraging globalized manufacturing is essential for large organizations to nimbly benefit from the rapid innovation of the microelectronics industry. However, its distributed nature allows for numerous points of attack throughout system lifecycles including production, deployment, operations, and support. One way to protect supply chains is by using advanced cryptographic hardware key storage solutions to store small amounts of private information on the protected item and permit its authentication throughout all points of attack in the chain. At these points, microelectronic components are vulnerable to tampering, substitution, and duplication of private key material stored in digital memory [1]. The high cost and complexity of current hardware key storage solutions make their practical scalability to low-end systems intractable [2], [3]. Moreover, the secure storage of private key material remains a challenge for all types of cryptographic systems [4].

Physical unclonable functions (PUFs) store private information in the intrinsic physical structure of a device and provide an alternative to the storage of key material in digital memory [5], [6]. These devices are most commonly applied to authentication applications [7]–[9] and more recently for secure key generation [10], [11]. While electronic PUFs remain the most common approach, many have been found susceptible to cloning, invasive interrogation, and model-building attacks [12]–[14]. Optics promises "strong PUFs" [15] through more complex physical interactions, and demonstrations thus far have leveraged inhomogeneous spatial scattering of light [7]–[11]. While these optical scattering approaches may offer additional security benefits and output complexity when compared to electronic PUFs [6], [10], they continue to be slow, large (~100 $mm^3$), extremely sensitive to mechanical positioning, and difficult to integrate with electronic circuits [7]–[11]. Further, as we show here, optical scattering PUFs provide a relatively low information density (storage of large amounts of cryptographic material in a small space) due to their linearity [7]–[11].

More precisely, optical scattering PUFs (OSPUFs) have exceptionally slow cryptographic key material ("information") generation rates due to requirements to mitigate induced heating from the input, the need for narrow linewidth laser illumination sources, and camera-based detection methods. Additionally, all demonstrated OSPUFs are linear devices, which adversely affects their security and their total information content. Linear OSPUFs can be attacked by machine learning techniques if the raw interference images of the PUF are available to the adversary [9]. Successful model-building attacks have been able to predict the correct response to a random and previously unobserved input challenge [9]. Also, due to their linear behavior the uncorrelated key count is limited to the total number of orthogonal input sequences, as such illumination patterns add linearly and coherently [10], thus reducing the total information content of the device. These non-ideal properties have precluded their practical use even after nearly two decades of research and development.

Recent work on ultrafast silicon micro-cavities has created an entirely new generation of optical PUFs that we term "photonic PUFs" [16], [17], and which overcome many of the challenges faced by OSPUFs. Driving unique, CMOS-compatible, silicon micro-cavities with spectro-temporally encoded, ultrashort optical pulses (Fig. 1) leads to the excitation of a large number of transient spatial optical modes [18] that mutually interact via the optical nonlinearities of silicon [19] and the precise physical structure of the cavity itself. Fabrication variances at the nanometer-scale are inevitable and form the device's unique information carrying structure. The cavity design is known to exhibit chaotic behavior [20] that is manifested in an extreme sensitivity to structural variations, thus rendering cloning virtually impossible [16]. Guided waves used to couple to and from the cavity require no spatial alignment, and the device is therefore significantly more robust than optical scattering approaches. Further, the device is readily integrated with silicon electronics, as it is constructed from the same material and fabrication process. By using a broadband source in the near-IR spectrum, this photonic PUF is compatible with modern telecommunications infrastructure and components. Lastly, the ultrafast response (~12 ps) denies an adversary sufficient time to emulate and spoof the device.

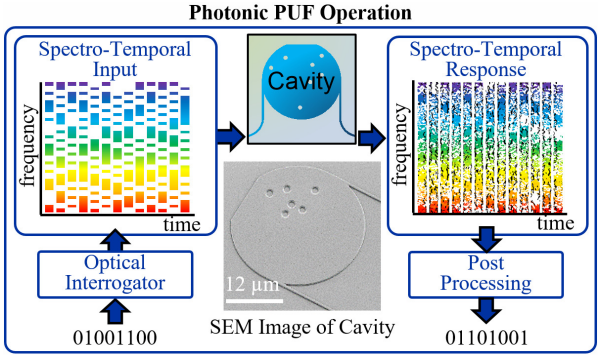

Fig. 1. Photonic PUF operation. A binary challenge sequence spectrally encodes a train of ultrashort optical pulses, exciting many spatial modes in the silicon micro-cavity and generating a unique spectro-temporal response that is converted into a binary sequence. Inset: Scanning electron microscope (SEM) image of prototype device.

Here we experimentally and theoretically investigate the information capacity (the number of random bits of information that it can store) of these novel photonic PUFs. Compared to existing optical PUFs [7]–[11], these devices demonstrate unprecedented information density and generation rates. This paper focuses on the impact of the device nonlinearity on the total information content. Consequently, we make the following new contributions:

- **Nonlinearity**: We exploit the natural nonlinearity of silicon to achieve enhanced security and resistance to model-building attacks. Our experimental results validate the nonlinear operation of the device.

- **Enhanced Information Density**: Nonlinearity affords substantial improvements in information density over OSPUFs, as the number of uncorrelated response patterns is bounded by the number of unique challenge patterns [8] up to a resolution set by the minimum feature size of the spectral response. These compact micro-cavities are much smaller than current OSPUFs, which is critical to enhancing information density.

- **Ultrafast Information Generation Rate**: Due to the ultrafast nature of the approach, we demonstrate continuous challenge-response measurements at a rate of 90-MHz yielding noteworthy improvements over the best OSPUFs demonstrated to date.

## II. NONLINEAR SILICON CHAOTIC CAVITIES

### A. Information Model

Experimentally we extract key information from the cavities in the following manner. A novel ultrafast pulse shaper spectrally encodes the amplitude of each laser pulse as follows [21]: Dispersion compensating fiber (DCF) is used to temporally disperse the individual pulses of a mode-locked laser (MLL) pulse stream. The temporally-dispersed spectra are then amplitude encoded by symbols of a pseudorandom binary sequence (PRBS) that is synchronized to the MLL. After spectral patterning, the pulses are compressed in time, amplified via an erbium-doped fiber amplifier (EDFA), and launched into the cavity where the complex nonlinear photonic interaction occurs. The optical pulses exiting the cavity are then pre-amplified for detection. The pulses are spectrally filtered to extract information from each response. The filtered pulse is detected and passed to an analog-to-digital convertor.

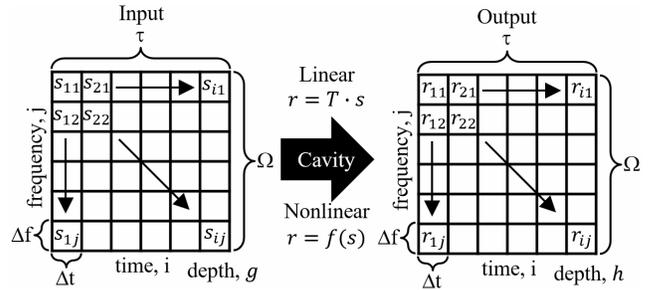

Fig. 2. Spectro-temporal input and output mapping. The device uses a time-frequency interrogation method across an input space bounded by the total spectral bandwidth and the cavity lifetime. These input symbols are mapped to time-frequency output symbols through the application of a transmission matrix in a linear system and an arbitrary function in a nonlinear system.

The input pulse is composed of multiple spectral input features of width $\Delta f$ and temporal features of duration $\Delta t$ that are bounded by (1) the total spectral bandwidth $\Omega$, which is limited by the bandwidth of the input MLL source, and (2) the total cavity lifetime, $\tau$, i.e. the time in which symbols may interact with each other (Fig. 2). These features are used to define the $i^{th}$ temporal and $j^{th}$ spectral feature of each symbol, $s_{ij}$. The total number of input time slots, $m_i$, is given by $m_i = \tau/\Delta t$, where $\Delta t$ is an arbitrary temporal feature size. Thus, it is desirable to maximize the cavity lifetime in order to gain the maximum interaction between input symbols. The cavity lifetime is influenced by the precise shape of the cavity, the material loss properties, and the refractive index contrast at the boundary between the cavity and surrounding cladding. The number of spectral features $m_j$ of an input is the total frequency bandwidth $\Omega$ divided by the input spectral feature size, $\Delta f_{in}$, i.e. $m_f = \Omega/\Delta f_{in}$. This input spectral feature size may be chosen arbitrarily but is limited by the modulation approach [21] and the subsequent correlation of excited responses (Section 3). In order to maximize the number of possible symbols, we can choose the time slot to be related to the inverse of the spectral feature, i.e. $\Delta f_{in} = 1/\Delta t$. The maximum number of input symbols is then $m = m_i m_j = (\Omega/\Delta f_{in})(\tau/\Delta t)$. By substituting this time-frequency relationship, we achieve $m = \Omega\tau$, or the total frequency bandwidth multiplied by the cavity lifetime. Each of these symbols can be encoded with $a$ bits of information via spectral amplitude and/or phase encoding for $g = 2^a$ levels.

Likewise, our detection approach is also spectro-temporal, i.e. energy may be measured in both time and frequency. The cavity lifetime $\tau$ and spectral bandwidth $\Omega$ are equivalent for both the input and output. We assume that each spectral response has a finite autocorrelation within some average spectral feature size $\Delta f_{out}$ and is zero outside of that bound. This estimate provides insight into the number of independent features contained within a response $p_f$ and is determined by dividing the total bandwidth $\Omega$ by the spectral feature size, i.e. $p_f = \Omega/\Delta f_{out}$. Each one of these features can be detected with a depth of $b$ bits (typically $b = 8$ bits for $h = 2^b$ levels).

Notably, there is no reduction in dimensionality versus three-dimensional OSPUFs, which are interrogated and read in a two-dimensional spatial basis as they are time-invariant. In contrast, our approach is dynamic in time and is thus three-dimensional (2D space and time) and is interrogated and read in a two-dimensional time-frequency basis as described above. Further, being spatially 2D (i.e. planar) is crucial for direct integration with semiconductor electronics.

*B. Nonlinearity, Information Content, and Security*

In a linear system, the mapping from input symbol $s$ to output symbol $r$ may be represented by a $m_i \times m_j$ transmission matrix, $T$ [9], [10]. Each output symbol may be represented as a linear combination of the input symbols by the relation $r = T \cdot s$. The maximum number of orthogonal rows of such a matrix is equivalent to its rank. An adversary can compute the inverse or pseudoinverse of $T$ to obtain, exactly or approximately, the input given the output. In a nonlinear system, the transmission function is a system of nonlinear equations for which no such inversion exists.

The property of nonlinearity is ideal for the construction of a PUF [7]–[9] as it breaks the linear dependencies that permit successful model building attacks and further enables the desired PUF properties of unpredictability and unclonability [2]. The majority of nonlinear systems remain impossible to solve analytically [22] as each such system is inherently coupled with aspects of itself, vastly complicating the search for an analytic representation of its behavior. Due to the susceptibility of OSPUFs to model-building attacks [9], finding suitable nonlinear materials for these devices remains a central open research problem which we do not address here. Further, the high input power levels that would be required to observe nonlinear behavior in current OSPUF materials are currently avoided due to the impracticality of such sources and the permanent device deformation that would fundamentally change the PUF response, rendering it unable to authenticate to previously enrolled authentication terminals [7], [11]. In contrast, our photonic PUFs readily exhibit nonlinear behavior due to the enhancement of the interaction time and optical intensity that accompanies the high optical confinement micro-resonant geometry and the naturally high nonlinearity of silicon [23]. To harness this nonlinearity, we excite a large number of long-lived optical cavity modes that nonlinearly mix with each other to maximize the spatial and spectro-temporal complexity of the interaction. To this end, silicon provides a high refractive index contrast with the silica cladding, thus improving cavity lifetime and confinement.

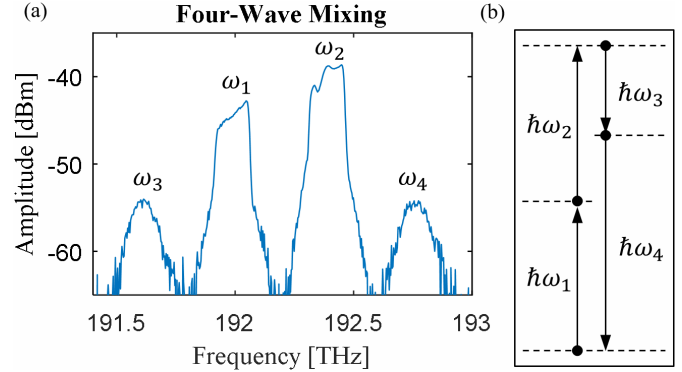

Fig. 3. Observed four-wave mixing in the photonic PUF. (a) An input signal consisting of two 150 GHz spectral features centered at $\omega_1 = 191.94$ THz and $\omega_2 = 192.43$ THz are amplified to an average power of 9.58 mW and sent through the silicon cavity. Optical power is observed at two new frequencies, $\omega_3 = 191.57$ THz and $\omega_4 = 192.80$ THz, as expected for this process. (b) Energy level diagram of FWM and optical frequency relationships.

Previous work [16] investigated the change in the output signal in both time and frequency as a function of power to characterize the nonlinear behavior, taking steps to ensure the invariance of the input laser spectrum. Such experimental measurements do not inform the precise source of the nonlinearity; however, several types are certainly present to varying degrees. In silicon, nonlinear effects are known to include self-phase modulation (SPM), two-photon absorption (TPA), cross-phase modulation (XPM), four-wave mixing (FWM), stimulated Raman scattering (SRS), and free-carrier induced absorption and dispersion. These effects and their power thresholds are well known and readily observed [23]. For example, we show the presence of FWM in a typical prototype device where two narrowband inputs nonlinearly interact to create light at new frequencies (Fig. 3).

## III. INFORMATION CONTENT OF PHOTONIC PUFs

*A. Information Content Estimation Approach*

In order to quantify the number of useful random bits, $N$, from the ideal photonic PUF device, we address the limiting physical phenomena [10]. First, correlations between features of each spectral response may limit the size of extractable information. Second, correlations among the set of possible spectral responses form an upper bound on the number of uncorrelated random responses from each device. The post-processing operations remove such correlations by reducing the number of output bits from each response [16]. We determine an approximate upper bound on the information content from the photonic PUF derived from experimental measurements, based upon the product of the number of random bits per spectral response $\beta$ and the total number of uncorrelated spectral responses $n$ per device [10]: $N \leq \beta \cdot n$.

*B. Bits per Spectral Response*

As our demonstrated system operates in a single time interval ($m_i = 1$) and uses a spectral filter to access several frequency components, the number of bits in a given response is the total spectral bandwidth, $\Omega$, divided by the feature size of the spectral response, $\Delta f_{out}$, multiplied by the bits per detected feature, $b$, i.e. $\beta \leq (\Omega/\Delta f_{out}) b$. First, we experimentally characterize our devices to determine the average feature size of the spectral output. We measure the input spectrum of a 300-fs full-wave half-maximum (FWHM) pulse train, send those pulses into a cavity at the transverse-electric (TE), transverse-magnetic (TM), and 45° cross-polarized (XP) polarization states, and measure the spectral response of 23 distinct cavities. We calculated $\Delta f_{out}$ as the FWHM of the averaged normalized autocorrelation function [24] resulting in 103.1±20.3 GHz for TE, 88.4±20.2 GHz for TM, and 46.9±24.2 GHz for XP (Fig. 4a). Hence, our rough estimate for an upper bound on the number of bits per response of our demonstrated 5-THz spectral bandwidth system is $\beta_{TE} \leq 388$ bits, $\beta_{TM} \leq 452$ bits, and $\beta_{XP} \leq 853$ bits.

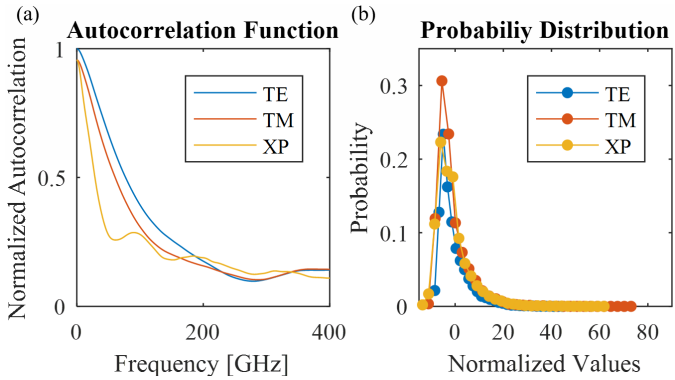

Fig. 4. Response characterization data. (a) Photonic PUF average feature size calculated as the FWHM of the normalized spectrum autocorrelation function for the TE, TM, and XP polarizations from 23 unique devices. (b) The spectral probability mass function is measured for the three polarizations.

Next, we refine the estimate of the number of random bits contained within each spectral response $\beta$ via calculating the entropy rate [10], $H(\nu)$, of the PUF spectral output $\nu$ from several devices. We first determine the spectral probability mass function from the measured spectral responses (Fig. 4b) normalized by mean and variance [24]. We then calculate $H(\nu)$ from the spectral probability mass function at our three polarizations to be 5.89 for TE, 5.97 for TM, and 5.42 for XP. We revise our estimate for the total random bits per spectral response as $\beta = (\Omega/\Delta f_{out}) H(\nu)$ resulting in $\beta_{TE} = 286$ bits, $\beta_{TM} = 338$ bits, and $\beta_{XP} = 578$ bits.

### C. Number of Uncorrelated Spectral Responses per Device

The number of uncorrelated spectral responses per device $n$ may be estimated experimentally. Due to the system's nonlinearity, the decorrelation of the resultant binary representation of the output is critical for determining the total number of uncorrelated responses. We compute this correlation via the fractional Hamming distance (FHD) between two binary sequences generated from unique spectral responses.

We experimentally characterize the correlation of the photonic PUF output spectra due to small changes to input spectra (Fig. 5a) at the TE polarization state. First, we generate a 67-feature frequency grid across the MLL input spectrum from 191.2-193.7 THz (2.5 THz) via a programmable spectral filter, yielding an input feature size of 37.3 GHz (emulating our authentication pulse shaping approach). Next, we generate a single pseudorandom binary input pattern and then 67 subsequent patterns each differing from the previous pattern by a single bit for 68 patterns in total. These patterns are sequentially encoded onto the broadband spectrum of the optical pulses, amplified, and measured before and after the cavity with an optical spectrum analyzer (OSA). In post-processing, we applied orthogonal, order-32, Hadamard sequences to channelize the output and extract spectral information (channel feature size of 78.1 GHz to be comparable with average feature size of the cavity) from each response. We then converted these channels into binary sequences by extracting five bits per channel (155 bits total). Due to the space limit, we will report more details of our post-processing approach in the full version of this paper.

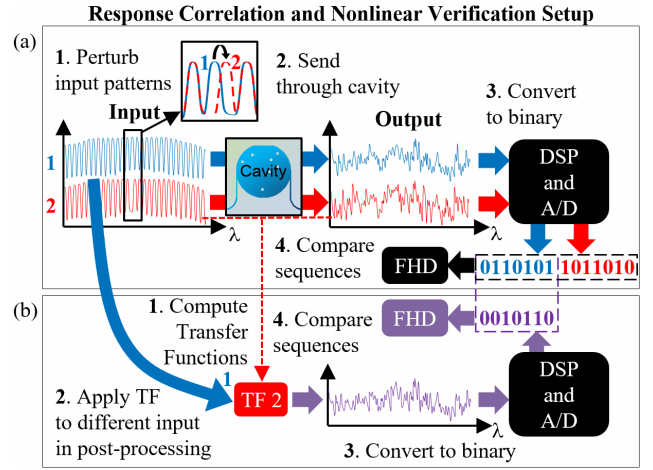

Fig. 5. Response correlation and nonlinear verification experimental setup. (a) Experimental characterization of the sensitivity of output binary sequences to small changes in the input spectra. Multiple input patterns were generated, differing by a single bit, and sent through the cavity at the TE polarization state. The responses were converted into binary sequences and compared via FHD. (b) Spectral transfer functions were computed for each unique input and output, applied to different input patterns in post-processing, and then converted to binary. The FHD was computed between binary sequences corresponding to different transfer functions applied to the same input pattern.

To ensure independence of binary responses from our demonstrated system, we evaluated the entropy of keys as follows (Fig. 6a). With an input source of 2.5 THz of bandwidth, we expect to extract $\beta = (2.5 \text{ THz} / 103.1 \text{ GHz}) \cdot 5.89 \approx 143$ bits per response. We calculate the entropy as 149 bits which is within the expected extractable information bound based upon the variance of our minimum feature measurements. We also characterized output binary response decorrelation as a function of deviation between input patterns (Fig. 6b). These results indicate that even a single bit difference between two input patterns will result in fully decorrelated binary responses from their corresponding outputs thereby satisfying the avalanche criterion [4].

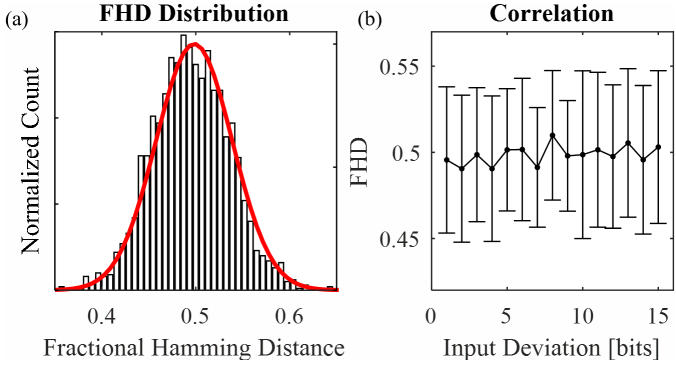

Fig. 6. Response correlation results. (a) FHD distribution between binary responses generated from PUF responses from deviated input patterns. The mean of this distribution is 0.498 and the standard deviation is 0.041, resulting in an entropy of [0.498×(1−0.498)]/0.041² =149 bits. (b) Correlation of output binary sequences by input pattern deviation (two standard deviation bounds).

In order to further verify the nonlinear behavior of the system, we first assume linear operation by calculating and analyzing the transfer functions corresponding to each pair of unique input and output spectra. In post-processing, we applied every combination of those transfer functions to each unique input to calculate 67 new outputs (Fig. 5b). We then converted each new output spectrum into a 155-bit sequence per our post-processing algorithm. We calculated the FHD of each reconstructed output with the originally measured output and determined the correlation between binary responses as a function of the deviation of the input pattern corresponding to the applied response (Fig. 7). The results indicate that output sequences based upon responses calculated from minutely varying inputs cause full decorrelation of the expected binary sequence thus showing that the behavior depends on the input sequence and that the device is verified to be nonlinear.

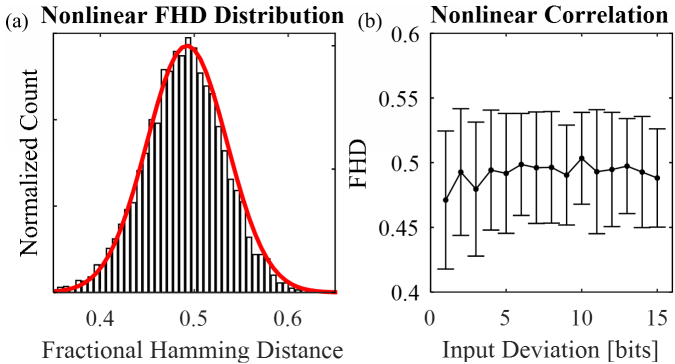

Fig. 7. Nonlinear verification results. (a) Nonlinear FHD distribution between binary responses generated from transfer functions from deviated input patterns. (b) Correlation of nonlinearly reconstructed binary sequences by input pattern deviation (two standard deviation bounds).

In the nonlinear regime, the responses do not add linearly or coherently, so the PUF can use many more challenges than in the linear case [7], [8]. Thus, the number of possible patterns forms an upper bound for information content computation and, the ideal upper bound on the total number of independent spectral responses is $n \leq g^{\Omega\tau}$. Our demonstrated ultrafast pulse shaper spectrally encodes the amplitude of each pulse to achieve 94 binary features within the 3-dB bandwidth (4.2 THz) of each pulse for $\Delta f_{in}$ = 44.2 GHz [21]. Likewise, we reduce the depth of the input to a single bit, i.e. $g$ = 2 levels, as our modulation approach is binary. Since our demonstrated system effectively integrates the response over time, we revise our upper bound to be limited by the number of input features across the spectral bandwidth, i.e. $n \leq 2^{\Omega/\Delta f_{in}} = 2^{94}$.

### D. Total Bits per Device and Ideal Upper Bound

Using $\beta$ and $n$ as computed above, we may now calculate an upper bound on the total number of random bits of our prototype photonic PUF as follows:

$$N_{TE} \leq \beta_{TE} \cdot n \approx 10^{22} \text{ Gbits} \quad (1)$$

For the device area of 707 μm², this yields a 2D information density of $10^{22}$ Tbits/mm². However, in the above analysis, the input optical bandwidth is limited by our experimental setup. Future work can employ a wider input bandwidth to extract much more information. While the low loss transmission window of silicon is 1.1 to 7 μm (229.9 THz), a practical system might use the near-infrared telecommunications band of 1.26 to 1.675 μm (59.0 THz). Also, we may further increase the information content through the use of orthogonal polarization states and by extending the input depth to more bits via phase and amplitude coding (e.g. $g = 2^4$ levels) such that the input space is increased. Thus, we can estimate a theoretical upper bound:

$$N \leq 2\Omega\tau \, H(\nu) \, g^{\Omega\tau} \approx 10^{847} \text{ Gbits} \quad (2)$$

This very optimistic upper limit suggests that future systems should continue to improve on information density and performance. While our experimental results (Figs. 5-7) suggest that an adversary cannot predict any new output from the observation of other input-output behavior, future research is needed to evaluate the extent to which the PUF continues to generate independent outputs in those conditions and therefore how close to this upper limit the PUF can operate.

### E. Performance Comparison

We now compare the results of our photonic PUF against various OSPUF demonstrations in key performance metrics of information generation rate, information content, and information density (Table 1). As OSPUFs continue to make use of linear materials and configurations, they continue to suffer from reduced information content. Our nonlinear photonic PUF provides an 18 order-of-magnitude improvement in information content due to the ability to leverage the entire challenge space. Further, given the large physical size of OSPUFs, in part due to requirements for free-space propagation, they are prohibitively difficult to integrate with electronic circuits and suffer from low information density. In contrast, our PUF is planar and is < 160 μm³ or six orders of magnitude smaller in volume, resulting in an improvement in information density of 27 orders of magnitude. Lastly, due to the ultrafast nature of the cavity, we can acquire [16] a measurement every 11 ns (seven orders of magnitude faster that previous OSPUF demonstrations) yielding ≥ 2 bits of key material with an overall key generation rate of over 180 Mb/s; a two order-of-magnitude improvement over the best OSPUFs demonstrated to date. Ultimately, the measurement speed is limited only by the minimum feature size, i.e. ~80 GHz features yielding a minimum of 12 ps per measurement.

TABLE I. OPTICAL PUF QUANTITATIVE PERFORMANCE COMPARISON

| PUF Type | PUF Performance Metrics | | | | |
|---|---|---|---|---|---|
| | Year | Info. Rate [bps] | Info. Limit [Gbits] | Volume [$mm^3$] | Info. Density [$Tbit/mm^3$] |
| OSPUF with Probe [7], [8] | 2002 | 233 | 5522 | 254 | 0.022 |
| OSPUF with SLM [10][a] | 2013 | $2 \times 10^6$ | 151 | 0.151 | 1 |
| Integrated OSPUF [9][b] | 2013 | $1.3 \times 10^5$ | ?[b] | 250[b] | ?[b] |
| Photonic PUF [16], [17] | 2016 | $1.8 \times 10^8$ | $10^{22}$ | $1.6 \times 10^{-7}$ | $6.3 \times 10^{25}$ |

[a.] Used for secure communications and encryption
[b.] Estimated/data not available

Beyond information content, the small size of the micro-cavities and their ultrafast nature contribute to the overall security of this approach. All previous PUFs claim that emulation is difficult; however, those claims are based purely qualitatively on the difficulty of building an emulation device. For an OSPUF, such a system would be required to respond in a fraction of a second. In comparison, to emulate our device, the system would need to begin responding in half of the cavity round-trip time or < 0.6 ps and complete its response in less than 12 ps to avoid detection. While construction of both emulation systems would be difficult, to do so with our approach is completely unforeseeable using any current processing or memory technology. For example, an adversary would be required to store at least half of a complete challenge-response library (CRL) of $10^{28}$ bits (10,000 yottabits) of information in an area of < 707 µm$^2$ so as not to violate the speed of information being bounded by the speed of light [25]. This lower bound density of $10^{13}$ Tbits/µm$^2$ is many orders of magnitude denser than any current 2D memory technology [26] and this assumes that other emulation components have zero latency, which is clearly not the case. Thus, we argue that our infeasibility of emulation is more tangible than any previous claim as it is based on strict physical limits.

IV. SUMMARY, DISCUSSION, AND FUTURE WORK

In this paper, we show that the nonlinear behavior of this new class of photonic PUFs is essential to achieve revolutionary advancements in total information content, information density, and information generation rates for optical PUFs. We envision such unspoofable keys directly integrated, at the point of manufacture, into the same silicon layer as the microelectronics. Due to the compact micron-scale footprint, keys can be incorporated into individual chips, ensuring authenticity and integrity at any subsequent stage. Finally, the keys are probed optically, allowing them to be either independent of the microelectronic circuit, or intrinsically linked via optoelectronic interfaces. Future work in this area will focus on continuing to improve the performance via the investigation of wider bandwidth sources, the use of rapid spectro-temporal multilevel amplitude and phase encoding, and temporal multiplexing to enhance key generation rates. We also envision the development of new designs that provide for longer photon lifetimes and increased nonlinearity, thus further increasing the information content.